\documentclass[12pt]{iopart}

\usepackage{graphicx}

\newcommand{\cfigl}[3]{\begin{figure}[!hbtp]\centering
 \includegraphics[width=.5\textwidth]{#2}\caption{\small{{#3}}}\label{#1}\end{figure}}
\newcommand{\dfiglcap}[4]{\begin{figure}[!hbtp]\centering
 \includegraphics[width=.35\textwidth]{#2}\hspace{0.5cm}\includegraphics[width=.35\textwidth]{#3}
\caption{\small{#4}}\label{#1}\end{figure}}
\usepackage{iopams}

\begin{document}

\title[Interaction Lagrangian for Dirac particles]
{An interaction Lagrangian for \\two spin $1/2$ elementary Dirac particles}

\author{Mart\'{\i}n Rivas}

\address{Theoretical Physics Department, The University of the Basque Country,\\ 
Apdo.~644, 48080 Bilbao, Spain}
\ead{martin.rivas@ehu.es}

\begin{abstract}
The kinematical formalism for describing spinning particles developped by the author
is based upon the idea that an elementary particle is a physical
system with no excited states. It can be annihilated by the interaction with its antiparticle
but, if not destroyed, its internal structure can never be modified.
All possible states of the particle
are just kinematical modifications of any one of them. 
The kinematical state space of the variational formalism of an 
elementary particle is necessarily a homogeneous space of the kinematical group of spacetime
symmetries. By assuming Poincar\'e invariance we have already described a model of a 
classical spinning particle which satisfies Dirac's equation when quantized.
We have recently shown that the spacetime symmetry group of this Dirac particle 
is larger than the Poincar\'e group. It also contains spacetime dilations
and local rotations. In this work we obtain
an interaction Lagrangian for two Dirac particles, which is invariant under this enlarged
spacetime group. It describes a short- and long-range interaction such that when averaged, 
to supress the spin content
of the particles, describes the instantaneous Coulomb interaction between them.
As an application, we analyse the interaction between two spinning particles, and show that
it is possible the existence of metastable bound states for two particles of the same charge,
when the spins are parallel and provided some initial conditions are fulfilled. 
The possibility of formation of bound pairs is due to the zitterbewegung spin structure of the particles 
because when the spin is neglected, the bound states vanish.
\end{abstract}


\vspace{1cm}
\noindent{\it J. Phys. A: Math. Theor.} {\bf 40} (2007) 1-12

\maketitle

\section{Introduction and scope}
We introduce in the next section a brief description of the kinematical formalism \cite{Rivasbook} for describing
classical elementary spinning particles. The kinematical group of spacetime
symmetries not only states the symmetries and conservation laws but also
provides the classical variables we need to describe 
an elementary particle. The Dirac particle, a classical system which satisfies Dirac's equation when quantized,
is summarized in section \ref{sec:Dirac}. It
is Poincar\'e invariant but the spacetime symmetry group has been 
recently enlarged \cite{Rivaspacetime} to include spacetime dilations and local rotations, 
so that the classical
Dirac particle has a larger kinematical group of symmetries. 
Classical electromagnetism usually analyses interactions between spinless 
test point particles.
If electromagnetism is a long range and also a short-range interaction it suggests 
that the same law applies at large and short distances
and therefore that a scale invariance is involved. 
We have models of spinning particles which have a spacetime dilation invariance
and we take advantage of this fact to analyse a long and a short range interaction 
among them and to check its behaviour when the spin is supressed.
In section \ref{sec:lag}, we construct a Lagrangian for describing the interaction between two Dirac particles
which is invariant under this enlarged group. Because it is also invariant
under spacetime dilations, it will describe a long and short-range interaction, which
suggests a kind of generalization of the instantaneous electromagnetic interaction,
because it has a Coulomb limit when the spin of the particles is smeared out.
We also discuss there its structure in a synchronous 
time description of the evolution. 
Section \ref{sec:int} is devoted to the analysis of the evolution of two particles of the same
charge when the spins are parallel to each other. 
One of the salient features is the possibility of formation,
from the classical point of view, of metastable bound states for 
Dirac particles of the same charge,
when some boundary conditions are fulfilled.

\section{Summary of the kinematical formalism}
\label{sec:form}
Because all known elementary particles, the quarks and leptons, are spinning particles
and it seems that there are no spinless elementary particles in nature,
we took the challenge of obtaining a classical formalism for describing spin.
The interest of a classical description of spinning matter is not important in itself, because
matter, at this level behaves according to the laws of quantum mechanics. But finer a
classical description of elementary matter a deeper quantum mechanical formalism,
because we will have at hand, when quantizing the system, 
more classical variables to deal with, and therefore
with a more clear physical and/or geometrical interpretation. A second feature
is that a classical formalism supplies models. Both goals, in my opinion, 
have been successfully achieved.

The kinematical formalism for describing elementary spinning particles \cite{Rivasbook},
previously aimed for the spin description of matter, has proven to be a general framework
for the description of elementary particles, because it supplies a very precise definition
of an elementary particle which has, as a quantum counterpart, Wigner's definition. All elementary
systems described within this formalism have the feature that when quantized their Hilbert space of pure states
carries a projective unitary irreducible representation of the kinematical group. It is through
Feynman's path integral approach that both formalisms complement each other.

It is based upon the four fundamental principles: Relativity principle, atomistic principle, variational principle
and uncertainty principle. The relativity principle states that there exists a set of equivalent observers, historically called
{\it inertial observers}, for whom the laws of physics must be the same. They are defined with respect to each other by
a spacetime transformation group, usually called the kinematical group of the formalism. The atomistic principle
admits that matter cannot be divided indefinitely. Matter does not statisfy the hypothesis of the continuum.
After a finite number of steps in the division of matter we reach an ultimate object, an {\it elementary particle}. 
The distinction between an elementary particle
and any other finite system is that an elementary particle has no excited states and, 
if not destroyed, it can never be modified, so that
all possible states are only kinematical modifications of any one of them.  This implies that the states
of an elementary can be described by a finite set of variables. The variational principle recognizes that the 
the action of the evolution of any mechanical system between some initial and final states must be stationary.
This completes the classical framework. For the quantum description we must susbtitute this last variational 
principle by the uncertainty principle, in the form proposed by Feynman: all paths of the evolution of any mechanical system 
between some initial and final states are equally probable. For each path a probability amplitude is defined,
which is a complex number of the same magnitude but whose phase is the action of the system between 
the end points along the corresponding path. In this way, classical
and quantum mechanics are described in terms of exactly the same set of classical variables.

This formalism determines that these variables, which define the initial and final states of the evolution
in the variational description, are a finite set of variables which necessarily span a 
homogeneous space of the kinematical group.
We call them the kinematical variables
of the particle. The manifold they span is larger than the configuration space and in addition
to the time and the independent degrees of freedom it also includes the derivatives of the 
independent degrees of freedom up to one order less the highest order they have in the Lagrangian.
The Lagrangian for describing these
systems will be thus dependent on these kinematical variables and their next order time derivative.
If the evolution is described in terms of some group invariant evolution parameter $\tau$, then,
when writting the Lagrangian not in terms of the independent degrees of freedom but as a function of 
the kinematical variables and their $\tau-$derivatives,
it becomes a homogeneous function of first degree of the $\tau- $derivatives
of all kinematical variables.

The formalism is completely general in the sense that it can accomodate any kinematical group we consider
as the spacetime symmetry group of the theory. 
But at the same time it is very restrictive, because once this 
kinematical group is fixed the kind of classical variables which define the initial and final states 
of an elementary particle in a variational approach, are restricted to belong to homogeneous spaces of the group. 
This kinematical group is the fundamental object of the formalism
and must be defined as a preliminary statement. 
We call the formalism {\it kinematical}, to stress this fact.
For the Galilei and Poincar\'e groups,
a general spinning elementary particle is just a localized and orientable mechanical system. 
By localized
we mean that to analyse its evolution in space we have just to describe the evolution of a single point ${\bi r}$,
where the charge is located and in terms of which the possible interactions are determined. 
This point  ${\bi r}$
also represents the centre-of-mass of the system for spinless particles, 
while for spinning ones must necessarily be a different
point than ${\bi q}$, the centre-of-mass, very well defined classically 
and where we can locate the mass of the particle. 
It is the motion of the charge around the centre
of mass which gives rise to a classical interpretation of the {\it zitterbewegung} and also to the
dipole structure of the particle.
By orientable we mean that in addition to the description of the evolution of the point charge we also need to
describe the change of orientation of the system by analyzing the evolution of a local comoving and rotating 
frame attached to that point. An elementary spinning particle is thus described as we use to describe a rigid
body but with some differences: we have not to talk about size or shape and the point does not represent
the centre-of-mass but rather the centre-of-charge. It is allowed to satisfy a fourth order 
differential equation and, for a Dirac particle, it moves at the speed of light. 

\section{A Dirac particle} 
\label{sec:Dirac}
For a relativistic particle the spacetime symmetry group is the Poincar\'e group.
In a recent work \cite{Rivaspacetime} we have shown that this spacetime symmetry group
for a classical spinning particle which satisfies Dirac's equation when quantized,
can be enlarged to include also spacetime dilations and local rotations of the body frame
associated to the particle. 
We denote this group for the classical Dirac particle by ${\cal S}={\cal W}\otimes SO(3)_L$, 
where ${\cal W}$ is the Weyl group, i.e., the Poincar\'e group ${\cal P}$ enlarged with spacetime dilations and
$SO(3)_L$ is the group of local rotations of the body frame, which commutes with ${\cal W}$.
The Lagrangian for a free Dirac particle is also invariant under this enlarged group ${\cal S}$.

If we consider this new group as the kinematical group of the theory, then the 
kinematical variables of a Dirac particle are reduced to time $t$, position of a point ${\bi r}$,
where the charge of the particle is located, its velocity ${\bi u}$ with the constraint
$u=c$, the orientation $\balpha$ which can be interpreted as the orientation of a local frame
with origin at point ${\bi r}$ and characterized by three parameters of a suitable parameterization
of the rotation group and, finally, a phase $\beta$ of the internal motion of the charge
around the centre-of-mass. If the particle has spin $S\neq0$ and mass $m\neq0$, then a length scale factor
$R=S/mc$ and a time scale factor $T=S/mc^2$ can be defined, such that all kinematical variables for the
variational description can be taken dimensionless. It is this argument which justifies the enlargement of
the spacetime symmetry group, to include spacetime dilations which preserve the speed of light.

The spinning particle has thus a centre-of-mass ${\bi q}$, which is always different than the point 
${\bi r}$, such that, for a free particle and for the centre-of-mass observer is at rest, 
and the point ${\bi r}$ is moving in circles
at the speed of light, around the point ${\bi q}$ in a flat trajectory contained in a plane
orthogonal to the spin ${\bi S}$. The kinematical variable $\beta$ describes the phase
of this internal or {\it zitterbewegung} motion.

The rotation subgroup of ${\cal P}$ transforms the kinematical variables ${\bi r}$, ${\bi u}$
and $\balpha$ among the inertial observers, and the local rotation subgroup $SO(3)_L$ only affects
to the local change of the particle frame, i.e., to the orientation variables $\balpha$.
Because the Weyl group has no central extensions \cite{Boya}, 
in the quantum representation the symmetry group becomes $\widetilde{\cal S}={\cal W}\otimes SU(2)_L$.

The general structure of the Lagrangian for a free Dirac particle when written in terms of the kinematical
variables and their $\tau-$derivatives is
\[
L=T\dot{t}+{\bi R}\cdot\dot{\bi r}+{\bi U}\cdot\dot{\bi u}+{\bi W}\cdot\bomega+B\dot{\beta}
\]
where $T={\partial L}/{\partial\dot{t}}$, ${\bi R}={\partial L}/{\partial\dot{\bi r}}$, 
${\bi U}={\partial L}/{\partial\dot{\bi u}}$, ${\bi W}={\partial L}/{\partial\bomega}$
and $B={\partial L}/{\partial\dot{\beta}}$, because of the homogeneity of $L$ in terms of the
$\tau-$derivatives of the kinematical variables.

\cfigl{fig:elec}{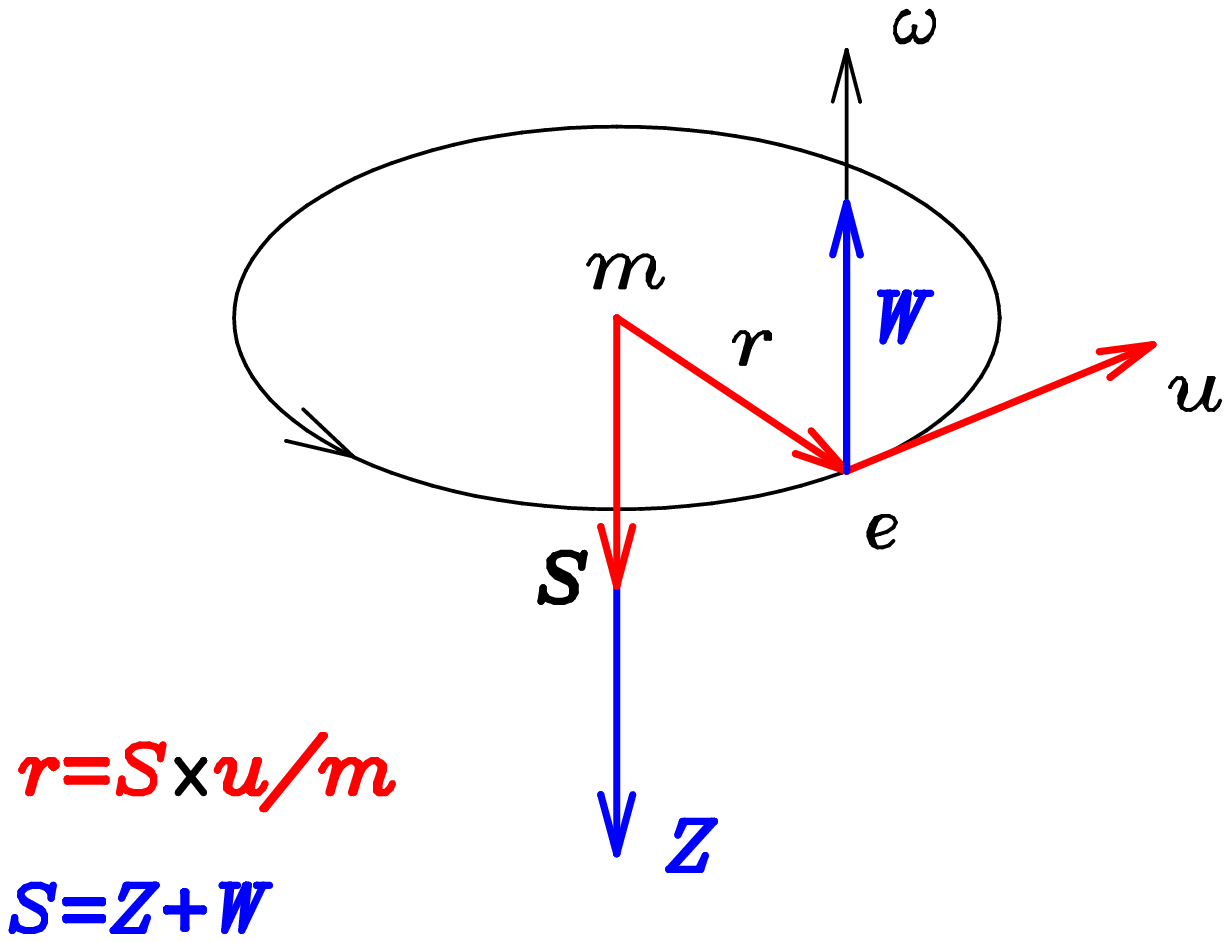}{Motion of the charge of a Dirac particle in the centre-of-mass frame. 
The total spin ${\bi S}$ is the sum of the orbital part ${\bi Z}$ and the rotational part of the body frame ${\bi W}$.
It is not depicted the local body frame, with origin at point ${\bi r}$, ${\bi e}_i$, $i=1,2,3$,
which rotates with angular velocity $\bomega$. The motion of the charge, with respect to the fixed spin
direction, is left-handed. The phase $\beta$ corresponds to the phase of the circular motion.}

The spin is defined by
\[
{\bi S}={\bi u}\times{\bi U}+{\bi W}\equiv {\bi Z}+{\bi W}
\]
where ${\bi Z}$ is the zitterbewegung part of spin and ${\bi W}$ is the rotational part as depicted in Fig.~\ref{fig:elec}.
The  ${\bi Z}$ part is related to the function  ${\bi U}$, i.e., to the dependence of the Lagrangian on the acceleration
$\dot{\bi u}$.
When quantizing the system the ${\bi Z}$ part of the spin quantizes with integer values while the spin $1/2$
is coming from the rotation part ${\bi W}$. It is this twofold structure of the spin which produces a clear interpretation
of the gyromagnetic ratio because the particle current, which produces the magnetic moment, is only related to the
zitterbewegung part of the spin ${\bi Z}$, which is twice the total spin ${\bi S}$, as seen in \cite{g2}.

The Casimir operators of the enlarged group ${\cal S}$ are the absolute value of the spin $S$, which is the Casimir
operator of the Weyl group ${\cal W}$ and the absolute value $I$ of the spin projection operator
on the body frame of the rotational part of the spin
\[
I_i={\bi e}_i\cdot{\bi W}
\]
which corresponds to the Casimir operator of the $SO(3)_L$ part. Here ${\bi e}_i$, $i=1,2,3$ represent
the three unit vectors of the local frame attached to the point ${\bi r}$.

A Dirac particle, with the enlarged group ${\cal S}$ as its kinematical group, has as intrinsic properties
the spin $S$ and the spin projection $I$ which take both the eigenvalue $1/2$ when quantized \cite{Rivaspacetime}. 
When quantizing the system only two kinds of Dirac particles arise according to the eigenvalues 
of the $Z$ part of the spin, which can take the values 0 or 1. The mass of the particle, very well 
defined as usual by means of the Casimir operator of the Poincar\'e group $H^2-{\bi P}^2$, and which must be different
from zero in order to properly define the Casimir  operator $S^2$, 
is no longer an intrinsic property of a Dirac particle, because this operator does not commute with
the generator $D$ of space time dilations. Mass, at this stage with a larger spacetime symmetry group, 
represents a property associated with the different states of the Dirac particle. It could mean, for instance,
that the electron and a massive neutrino are two different states of the same particle, the lepton.

It is shown that the dynamical equation of point ${\bi r}$ for the free particle and 
in the centre-of-mass frame is given by
 \begin{equation}
{\bi r}=\frac{1}{mc^2}{\bi S}\times{\bi u},
 \label{eq:dyq}
 \end{equation}
and where the spin vector ${\bi S}$ is constant in this frame, as depicted in Fig.~\ref{fig:elec}.
The average value of the position $<{\bi r}>={\bi q}$ is the centre-of-mass position, which is zero in this frame.
The average value of the velocity $<{\bi u}>={\bi v}$ represents the centre-of-mass velocity,
which is also zero in this frame.
The radius of the zitterbewegung motion is $R=S/mc$, which becomes $\hbar/2mc$, when quantized,
i.e., half Compton's wavelength. 
This dynamical equation (\ref{eq:dyq}) is independent of the particular free Lagrangian we take provided the kinematical variables
remain the same as the ones quoted here.

\section{An interaction Lagrangian}
\label{sec:lag}

An elementary particle can be annihilated by the interaction with the corresponding antiparticle, 
but if it is not destroyed, we made the assumption that the structure of an elementary particle 
is not modified by any interaction so that its intrinsic properties, the spin $S$ and the spin projection
on the body frame $I$ cannot be altered by the interaction with an external field
or by the presence in its neigbourhood of any other particle. 

Let us consider a compound system formed by two spinning particles with the same kind of kinematical
variables. We shall use a subscript $a=1,2$ to distinguish the variables corresponding
to each particle. Then the kinematical space of the compound system is spanned by the variables
$(t_a,{\bi r}_a,{\bi u}_a,\balpha_a,\beta_a), a=1,2$. The Lagrangian of the system will be written
as
\[
L=L_1+L_2+L_I
\]
where the $L_a$, $a=1,2$, are the free Lagrangians of each particle and $L_I$ is the interaction
Lagrangian we are looking for. Both $L_a$ are invariant under the enlarged group ${\cal S}$ and we are going
to find an interaction Lagrangian $L_I$ also invariant under ${\cal S}$.
The general structure of the free Lagrangian $L_a$ of each particle, which only depends on the corresponding 
kinematical variables of particle $a$, is
\[
L_a=T_a\dot{t}_a+{\bi R}_a\cdot\dot{\bi r}_a+{\bi U}_a\cdot\dot{\bi u}_a+{\bi W}_a\cdot\bomega_a+B_a\dot{\beta}_a
\]
where $T_a={\partial L_a}/{\partial\dot{t}_a}$, ${\bi R}_a={\partial L_a}/{\partial\dot{\bi r}_a}$, 
${\bi U}_a={\partial L_a}/{\partial\dot{\bi u}_a}$, ${\bi W}_a={\partial L_a}/{\partial\bomega_a}$
and $B_a={\partial L_a}/{\partial\dot{\beta}_a}$, because of the homogeneity of each $L_a$ in terms of the
$\tau-$derivatives of the corresponding kinematical variables.
The spin and the spin projection on the body frame for each particle, are
\[
{\bi S}_a={\bi u}_a\times{\bi U}_a+{\bi W}_a,\quad {I_a}_i={\bi e_a}_i\cdot{\bi W}_a
\] 
where ${\bi e_a}_i$, $i=1,2,3$ are three orthogonal unit vectors with origin at point ${\bi r}_a$.

The interaction Lagrangian between 
these two particles $L_I$ will be in general a function of the kinematical variables of both particles
and of their $\tau-$derivatives.
If both intrinsic properties $S_a$ and $I_a$ of each particle are not modified by any interaction 
then the interaction Lagrangian cannot be a function of the derivatives of the kinematical 
variables $\dot{\bi u}_a$ and $\bomega_a$, $a=1,2$. Otherwise the functions ${\bi U}_a$ and ${\bi W}_a$
will be different than in the free case.
In this case the functions  ${\bi U}_a$ and ${\bi W}_a$, which give rise to the definition
of the spin, are obtained only from the corresponding free Lagrangian $L_a$.

Then, as far as the $\tau-$derivatives of the kinematical variables are concerned, the interaction 
Lagrangian $L_I$
will only depend on the variables $\dot{t}_a$, $\dot{\bi r}_a$ and $\dot{\beta}_a$, $a=1,2$. 
In addition to this,
it will also be a function of the kinematical variables $t_a$, ${\bi r}_a$, ${\bi u}_a$ and $\beta_a$, but not
of $\balpha_a$ because of the invariance under the local rotation group $SO(3)_L$. 
Spacetime dilation invariance implies that the Lagrangian is a function of the phase difference $\beta_1-\beta_2$,
and of $\dot{\beta}_1-\dot{\beta}_2$, but being both phases completely arbitrary and independent of each other
it means that must be independent of these variables.

Because of the constraint ${\bi u}_a=\dot{\bi r}_a/\dot{t}_a$, the interaction Lagrangian 
will thus be finally a function
\[
L_I=L_I(t_a,{\bi r}_a,\dot{t}_a,\dot{\bi r}_a),
\]
and a homogeneous function of first degree of the derivatives $\dot{t}_a,\dot{\bi r}_a$,
$a=1,2$. 

If we call as usual the Minkowski four-vector $x^\mu_a\equiv(t_a,{\bi r}_a)$, translation invariance implies that the Lagrangian must be a function
of $x^\mu_1-x^\mu_2$. The following two
terms $\eta_{\mu\nu}\dot{x}^\mu_1\dot{x}^\nu_2$ and $\eta_{\mu\nu}(x_1^\mu-x_2^\mu)(x_2^\nu-x_1^\nu)$,
where $\eta_{\mu\nu}$ is Minkowski's metric tensor, are Poincar\'e invariant. 
If we consider that the evolution parameter
$\tau$ is dimensionless, these terms have both dimensions of length squared. It therefore implies that its quotient is 
dimensionless and therefore invariant under spacetime dilations. 
Another requirement is that the Lagrangian is necesarilly a homogeneous function of first
degree in the $\tau-$derivatives of the kinematical variables. The squared root will do the job.
An additional discrete symmetry will be assumed because when the two particles are the same, and therefore indistinguishable,
the interaction Lagrangian must be invariant under the interchange $1\leftrightarrow 2$ between the labels
of the two particles.
We thus arrive at the ${\cal S}$ group invariant 
Lagrangian
 \begin{equation}
L_I=g\sqrt{\frac{\eta_{\mu\nu}\dot{x}^\mu_1\dot{x}^\nu_2}{\eta_{\mu\nu}(x_1^\mu-x_2^\mu)(x_2^\nu-x_1^\nu)}}=
g\sqrt{\frac{\dot{t}_1\dot{t}_2-\dot{\bi r}_1\cdot\dot{\bi r}_2}{({\bi r}_2-{\bi r}_1)^2-(t_2-t_1)^2}}
 \label{eq:LAgbuena}
 \end{equation}
where $g$ is a coupling constant. 
This Lagrangian describes an interaction which is scale invariant and thus
it is valid as a long and short range interaction and which 
has a Coulomb-like behaviour when the spin is supressed, as we shall see in the next section. 
In this way it suplies a kind of generalization of an action at a distance electromagnetic interaction
between spinning particles. 

Alternative Lagrangians which fulfil the requirements of invariance and homogeneity can be constructed.
For instance, $L=g(\dot{x}_1-\dot{x}_2)^\mu(x_1-x_2)_\mu/(x_1-x_2)^2$, 
but this one is a total $\tau-$derivative of the function $\log(x_1-x_2)^2$. 
Another could be $L=g(\dot{x}_1+\dot{x}_2)^\mu(x_1-x_2)_\mu/(x_1-x_2)^2$, also dimensionless and linear 
in the derivatives of the kinematical variables,
but it reverses its sign under the interchange $1\leftrightarrow 2$, and thus all interaction observables, like the interaction
energy are reversed, which is physically meaningless for two alike particles. 

The interaction Lagrangians $L=g\dot{x}_1^\mu\dot{x}_{2\mu}/(\dot{x}_1-\dot{x}_2)^\mu(x_1-x_2)_\mu$
and $L=g|(\dot{x}_1+\dot{x}_2)^\mu(x_1-x_2)_\mu|/(x_1-x_2)^2$ 
also fulfil the requirements of being invariant under 
the enlarged group ${\cal S}$, homogeneous of first degree in the 
derivatives of the kinematical variables, and invariant under the interchange of the particle labels, 
but they do not have a Coulomb-like limit when the spin is smeared out. They will not be considered here
and the analysis of the interaction they describe is left to a subsequent research. 
Perhaps many other Lagrangians could be guessed
according to the above requirements but we have not succeded in finding another one with such a desirable
Coulomb-like behaviour.

The interaction between two Dirac particles is not unique. We know that among leptons and quarks there are at least,
short range interactions like the weak and strong interactions and a short and long range one like the electromagnetic interaction.
The novelty is that this classical interaction Lagrangian, which possesses a Coulomb limit, 
describes the interaction between two spinning particles, which satisfy Dirac's equation when quantized,
and has been obtained by assuming a spacetime symmetry group larger than the Poincar\'e group. 

\subsection{Synchronous description}

Once an inertial observer is fixed we shall consider a synchronous time description, i.e.
to use as evolution parameter the own observer's time $t$ 
which is the same as the two time variables $t_1$ and $t_2$. In this case, 
$t=t_1=t_2$, $\dot{t}_1=\dot{t}_2=1$, and thus
 \begin{equation}
L_I=g\sqrt{\frac{1-{\bi u}_1\cdot{\bi u}_2}{({\bi r}_2-{\bi r}_1)^2}}=g\frac{\sqrt{1-{\bi u}_1\cdot{\bi u}_2}}{r}
 \label{eq:li}
 \end{equation}
where $r=|{\bi r}_1-{\bi r}_2|$ is the instantaneous separation between the corresponding charges
in this frame and ${\bi u}_a=d{\bi r}_a/dt$ the velocity of the charge of particle $a$.

An average over the charge position and velocity in the centre-of-mass
of particle 1 implies that $<{\bi r}_1>={\bi q}_1$ and $<{\bi u}_1>=0$, so 
that the interaction becomes the instantaneous
Coulomb interaction, between the centre-of-mass of the first particle 
and the charge position of the other. 
The average over the other then corresponds to the instantaneous Coulomb interaction of 
two spinless point particles because when neglecting the zitterbewegung we are
suppressing the spin structure.

It is suggesting that $g\sim \pm e^2$ in terms of the electric charge of each particle and where the $\pm$ sign
depends on the kind of particles either of opposite or equal charge.

\section{Analysis of a two-particle system}
\label{sec:int}

The dynamical equation of a free Dirac particle and for any inertial observer 
is a fourth-order differential equation for the position of the
charge ${\bi r}$ which can be separated into a system of coupled second order differential equations for the centre-of-mass ${\bi q}$
and centre-of-charge ${\bi r}$ in the form:\cite{dyn}
\[
\ddot{\bi q}=0,\quad \ddot{\bi r}=\frac{1-\dot{\bi q}\cdot\dot{\bi r}}{({\bi q}-{\bi r})^2}({\bi q}-{\bi r}),
\]
where now the dot means time derivative. The first equation represents the free motion of the centre-of-mass
and the second a kind of relativistic harmonic oscillation of point ${\bi r}$ around point ${\bi q}$ which preserves
the constant absolute value $c$ of the velocity $\dot{\bi r}$. In fact, if $\dot{q}\ll\dot{r}=1$, $|{\bi q}-{\bi r}|\sim1$
and the equation is just the harmonic motion $\ddot{\bi r}+{\bi r}\simeq{\bi q}$, of point ${\bi r}$ around ${\bi q}$. The 
factor $({1-\dot{\bi q}\cdot\dot{\bi r}})/{({\bi q}-{\bi r})^2}$ prevents that when we take the boundary value 
$\dot{r}(0)=1$, the solution does not modify this absolute value of the velocity of the charge.

In the case of interaction this second equation remains the same because it corresponds to the definition
of the centre-of-mass position which is unchanged by the interaction, because it only involves the ${\bi U}$ and ${\bi W}$
functions. 
The first equation for particle $a$ is going to be replaced by $d{\bi p}_a/dt={\bi F}_a$ where ${\bi p}_a$
is the corresponding linear momentum of each particle expressed as usual in terms of the centre-of-mass velocity 
\[
{\bi p}_a=\gamma({\dot{\bi q}_a})m\dot{\bi q}_a,\qquad \gamma({\dot{\bi q}_a})=(1-{\dot{\bi q}_a}^2)^{-1/2},
\]
and the force ${\bi F}_a$ is computed from the interaction Lagrangian (\ref{eq:li})
\[
{\bi F}_a=\frac{\partial L_I}{\partial {\bi r}_a}-\frac{d}{dt}\left(\frac{\partial L_I}{\partial{\bi u}_a}\right)
\]
For particle $1$ it takes the form:
 \begin{equation}
{\bi F}_1=-g\frac{{\bi r}_1-{\bi r}_2}{|{\bi r}_1-{\bi r}_2|^3}\sqrt{1-{\bi u}_1\cdot{\bi u}_2}+
\frac{d}{dt}\left(\frac{g{\bi u}_2}{2|{\bi r}_1-{\bi r}_2|\sqrt{1-{\bi u}_1\cdot{\bi u}_2}}\right)
 \label{eq:F}
 \end{equation}
where it contains velocity terms which behave like $1/r^2$ and acceleration terms which go as $1/r$
in terms of the separation of the charges $r=|{\bi r}_1-{\bi r}_2|$. In this new notation ${\bi u}_a=\dot{\bi r}_a$.

Then the system of second order differential equations to be solved are
 \begin{eqnarray}
\ddot{\bi q}_a&=&\frac{\alpha}{\gamma({\dot{\bi q}_a})}\left({\bi F}_a-\dot{\bi q}_a({\bi F}_a\cdot\dot{\bi q}_a)\right)\label{eq:q2}\\
\ddot{\bi r}_a&=&\frac{1-\dot{\bi q}_a\cdot\dot{\bi r}_a}{({\bi q}_a-{\bi r}_a)^2}({\bi q}_a-{\bi r}_a),\quad a=1,2\label{eq:r2}
 \end{eqnarray}
where $\alpha$ is the fine structure constant once all the variables
are taken dimensionless. For that, we take the space scale factor $R=\hbar/2mc$ and the time scale as $T=\hbar/2mc^2$. 
All terms of equation (\ref{eq:q2})
which depend on the acceleration of the charges have to be replaced by the expressions of (\ref{eq:r2}).

It would be desirable to find analytical solutions of the above equations (\ref{eq:q2}-\ref{eq:r2}). Nevertheless 
we have not succeded in finding such a goal. 
However we shall analyse different solutions obtained by numerical integration.
We are going to use the computer program Dynamics Solver \cite{JMA}. The quality of the numerical 
results is tested by using the different integration schemes this 
program allows, ranging from the very stable embedded Runge-Kutta code 
of eighth order, due to Dormand and Prince, to very fast extrapolation 
routines. All codes have adaptive step size control and we check that 
smaller tolerances do not change the results. Another advantage is that it can be prepared
to analyse solutions corresponding to a wide range of boundary conditions, automatically.

\cfigl{fig:scat}{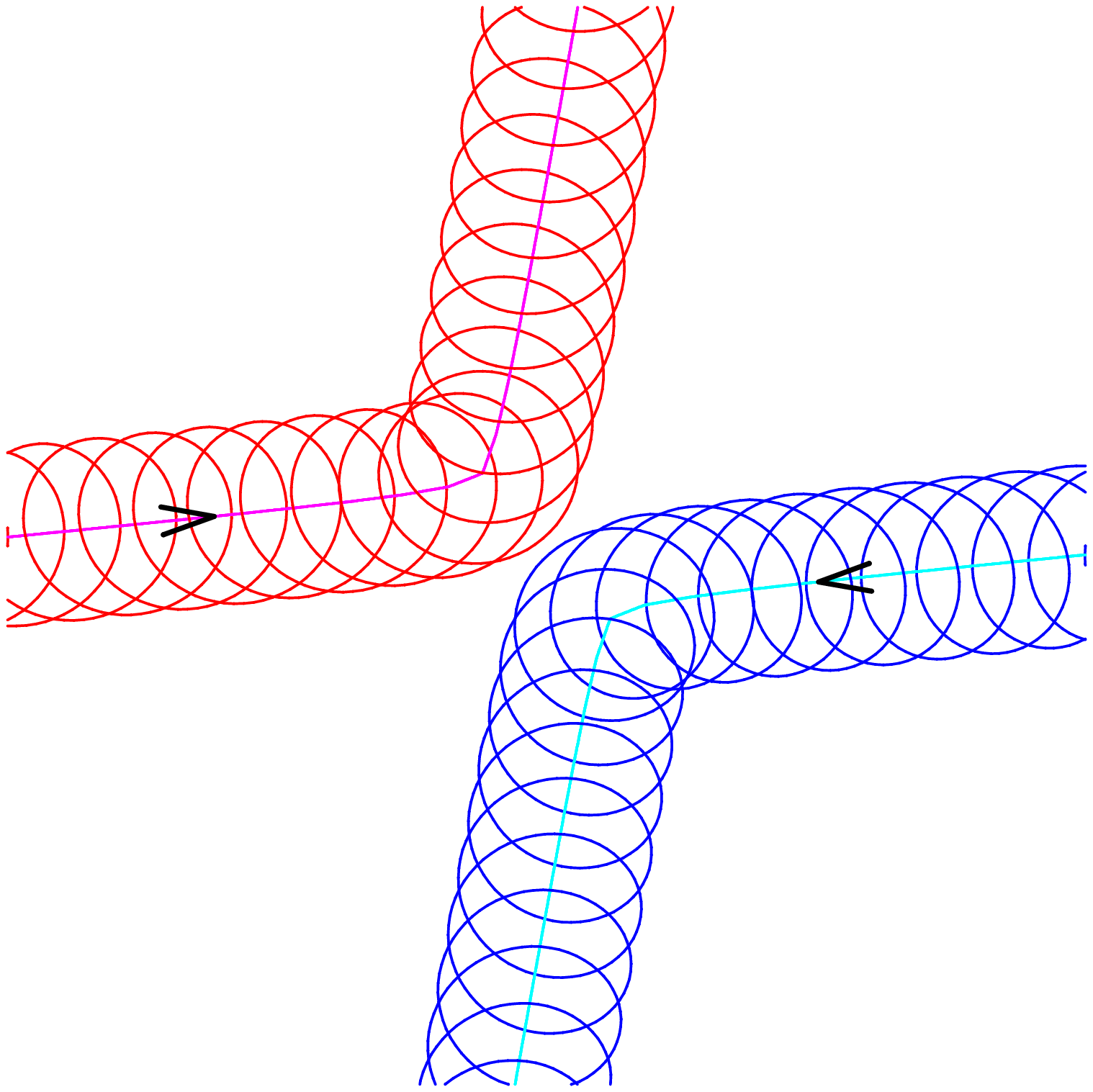}{The trajectories of the centres of mass and charge of two spinning particles 
with an initial centre-of-mass velocity $\dot{q}_a=0.1$ and a small impact parameter.}

See in figure \ref{fig:scat} the scattering of two equal charged particles with parallel spins.
The centre-of-mass motion of each particle is depicted with an arrow. 
If the two particles do not approach
each other too much these trajectories correspond basically to the trajectories of two spinless point particles
interacting through an instantaneous Coulomb force. By too much we mean that their relative separation
between the corresponding centres of mass
is always much greater than Compton's wavelength. This can be understood because
of the above discussion about the Coulomb behaviour of the averaged interaction Lagrangian, 
if the average position of each charge is far from the other. 
For high energy interaction the two particles approach
each other below that separation and therefore the average analysis no longer works because
the charges approach each other to very small distances where the interaction 
term and the exact position of both charges, becomes important. 
In this case new phenomena appear. We can have, for instance, a forward scattering like the one depicted 
on figure \ref{fig:forw}, which is not described in the classical spinless case, 
or even the formation of bound pairs for particles of 
the same charge, which we shall analyse in what follows.

\cfigl{fig:forw}{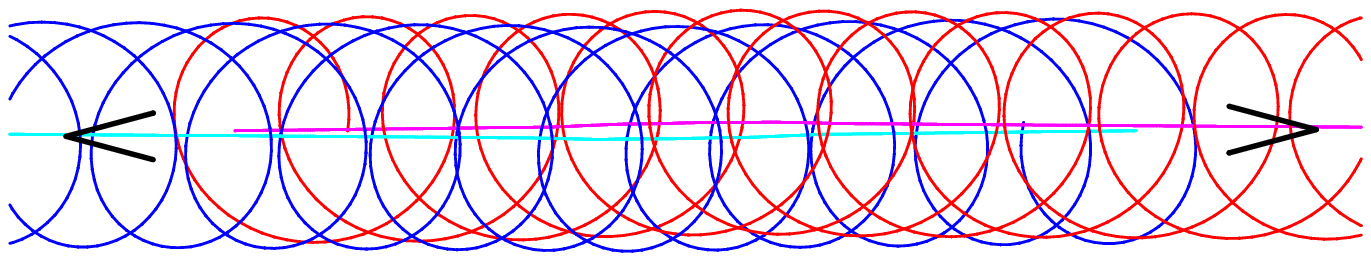}{Forward scattering of two spinning particles of the same charge 
with an initial separation $2q_a(0)=10$, centre-of-mass velocity $|\dot{q}_a(0)|=0.18$ and a very small impact parameter. The two centres of mass 
cross very close to each other, with a small deviation.}

In figure \ref{fig:phase} we represent an initial situation 
for two equal charged particles with parallel spins such that the corresponding centres of mass are separated
by a distance below Compton's wavelength. 
Remember that the radius of the internal motion is half Compton's wavelength.
We locate the charge labels $e_a$ at the corresponding points ${\bi r}_a$
and the corresponding mass labels $m_a$ to the respective centre-of-mass ${\bi q}_a$.
We depict in part (a) the situation when the two particles have the same phase $\beta_1=\beta_2$.
The forces ${\bi F}_a$, on each particle $a=1,2$, are  computed in terms of the positions,
velocities and accelerations of both charges,  according to (\ref{eq:F}),
and are also depicted on the corresponding centres of mass as a consequence of the structure of 
the equations (\ref{eq:q2}). We see that a repulsive force between the charges
produces also a repulsive force between the centres of mass in this situation. However, 
in part (b) both charges have opposite phases $\beta_1=-\beta_2$, 
and now the repulsive force between the charges
implies an attractive force between the corresponding centres of mass. 
If the initial situation is such that the centres-of-mass
separation are greater than Compton's wavelength, the force is always repulsive 
irrespective of the internal phases of the particles.

\dfiglcap{fig:phase}{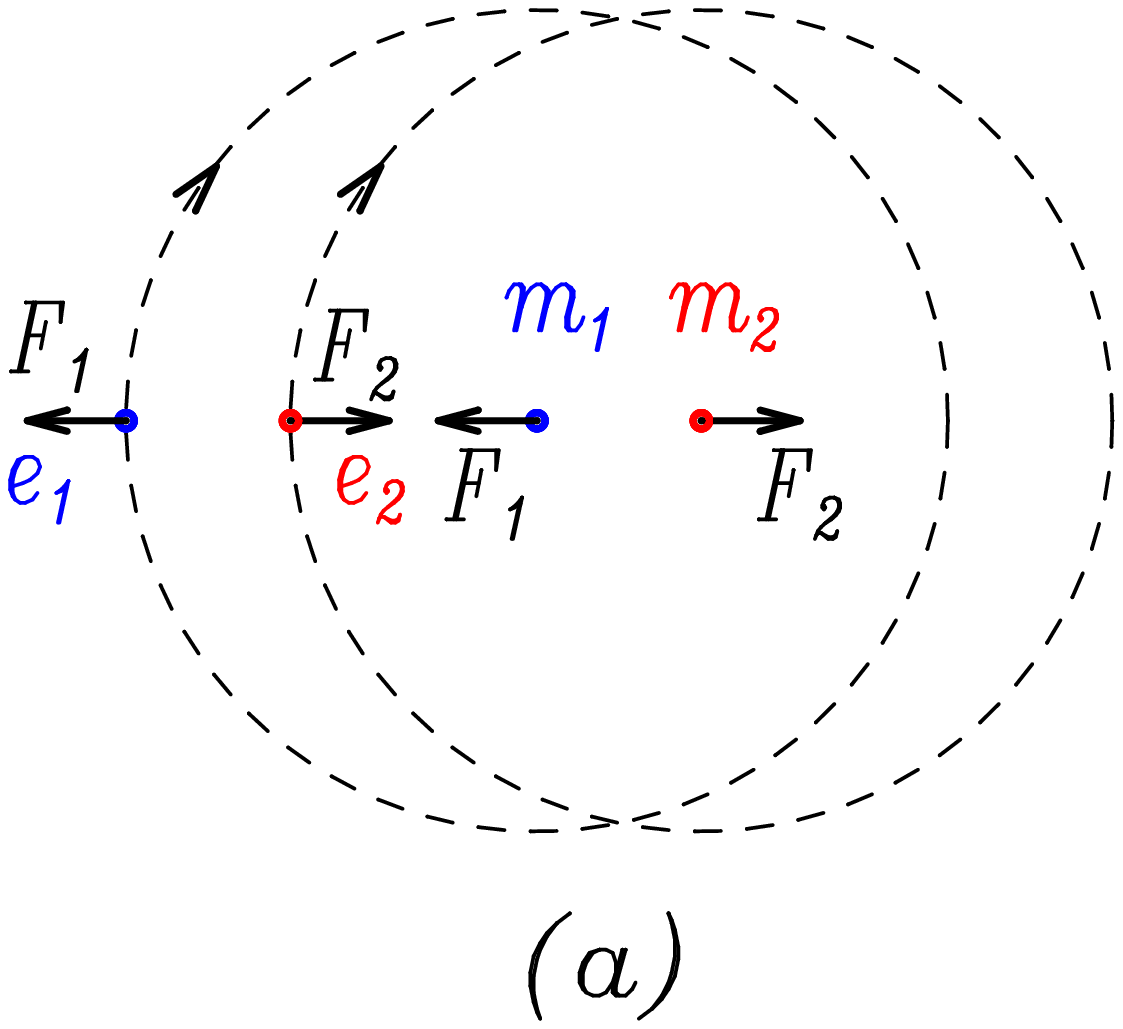}{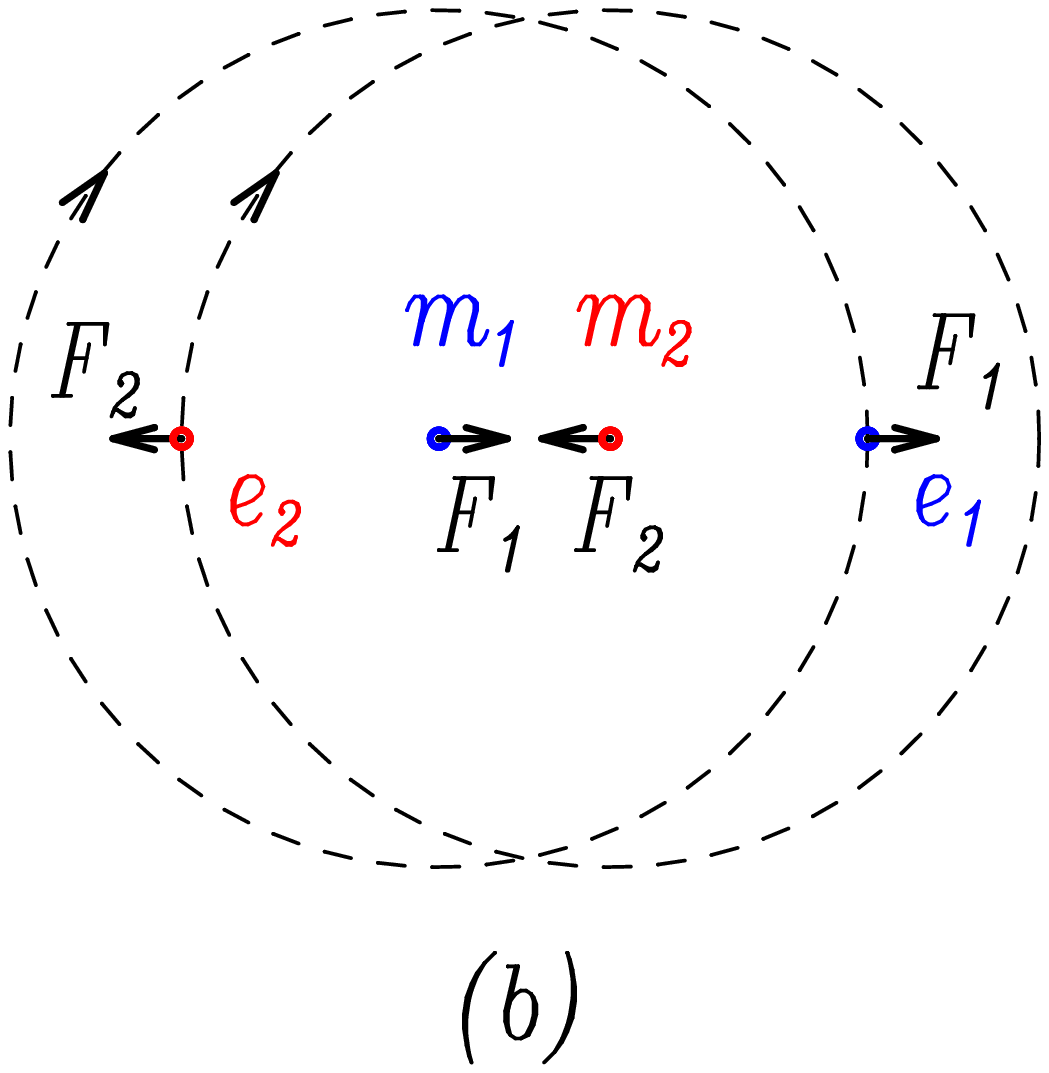}{Boundary values for two Dirac particles with parallel spins and with a separation between the centres
of mass below Compton's wavelength. The dotted lines represent the previsible clockwise motion of each charge. 
In (a) both particles have the same phase and the repulsive force between charges
produces a repulsive force between their centres of mass, while in (b), with opposite phases, the force between
the centres of mass is atractive.}

In figure \ref{fig:phaseL} we have another situation of opposite phases and where the initial separation between the centres of mass is larger
but still smaller than Compton's wavelength.
\dfiglcap{fig:phaseL}{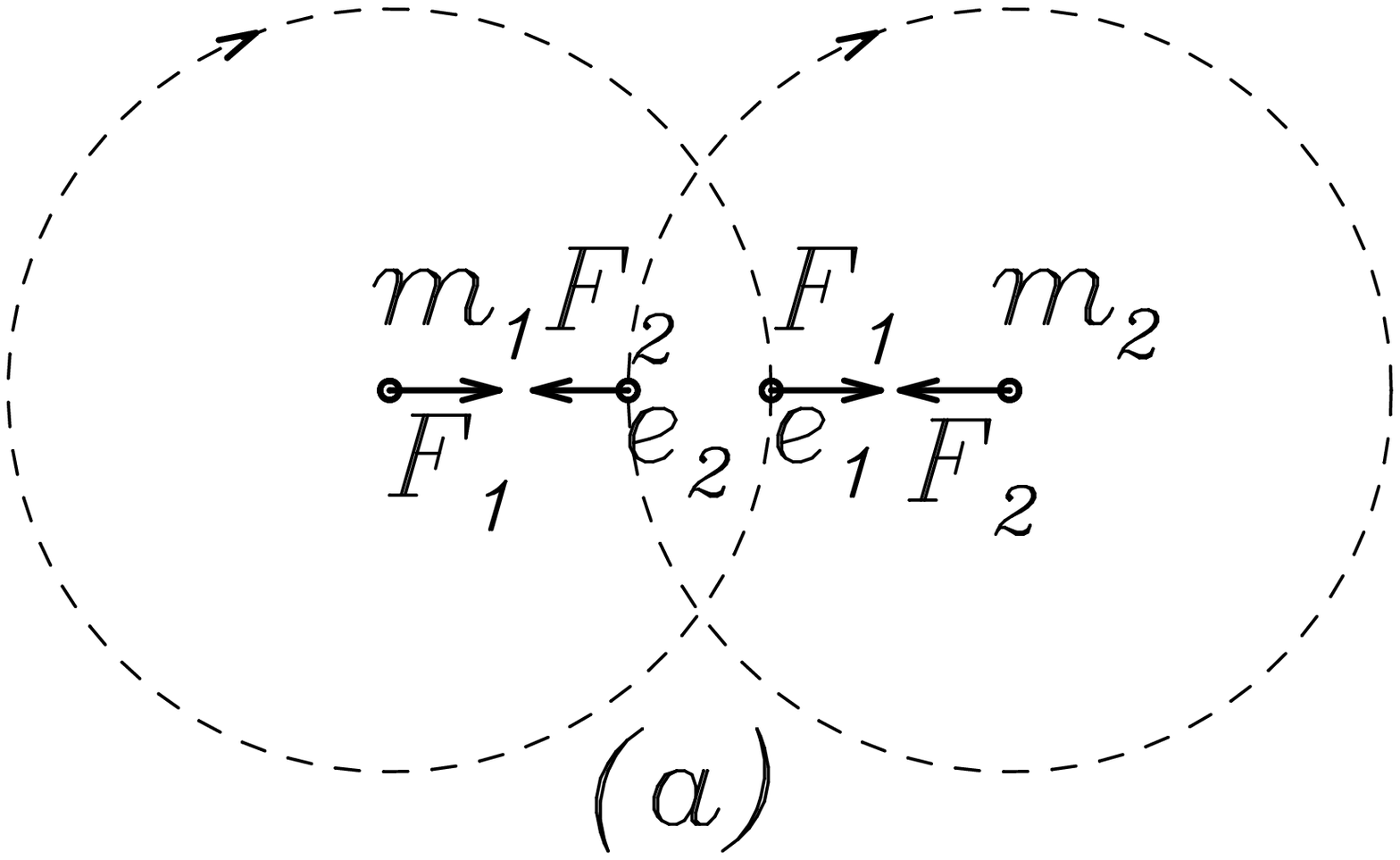}{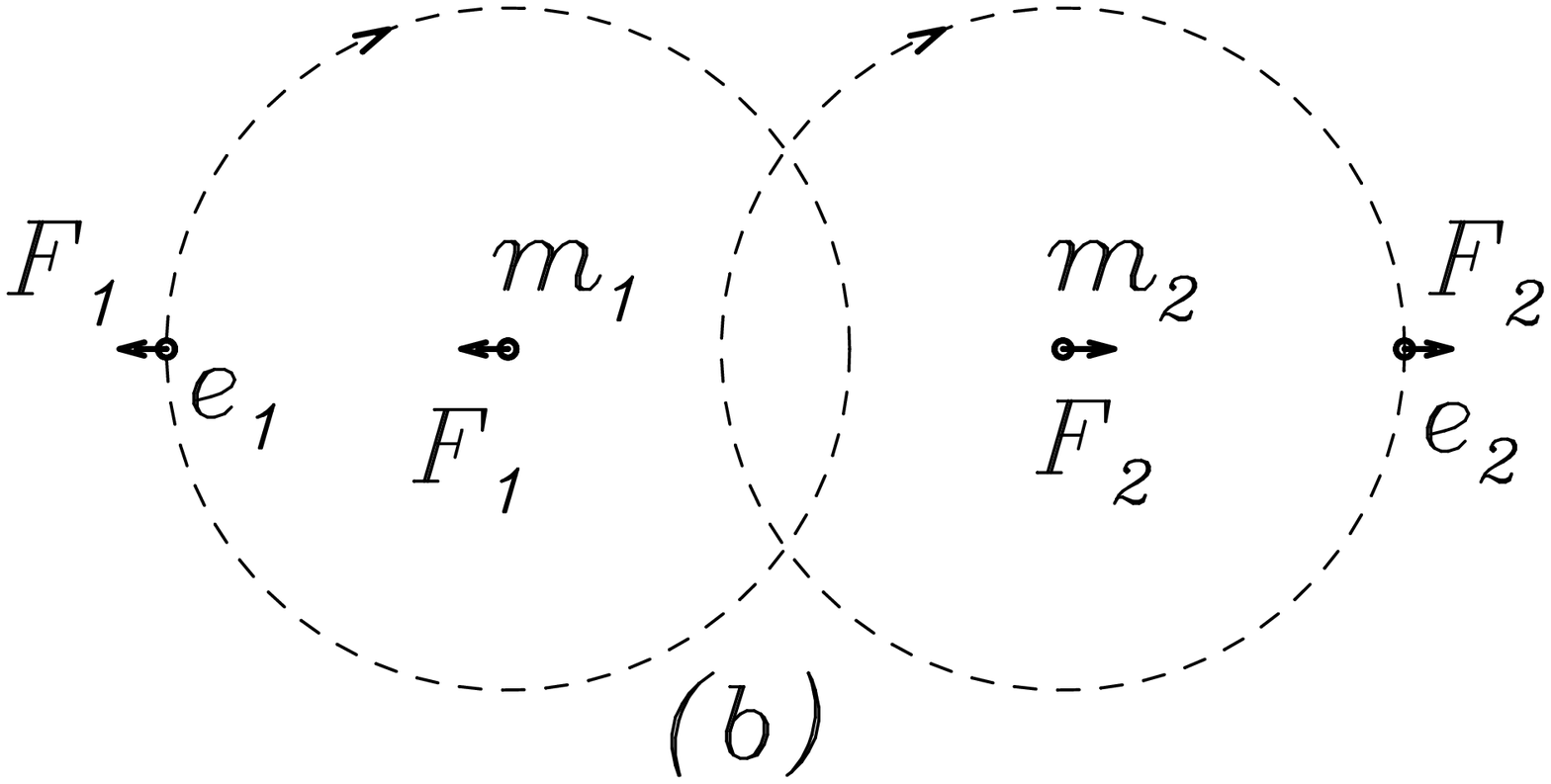}{(a) Another situation of two charges with opposite phases which produce an atractive
force between the centres of mass provided they are separated below Compton's wavelength. In part (b), after half a 
cycle of the motion of the charges, the force becomes repulsive between the centres of mass, but its intensity
is much smaller than the atractive force in (a) so that the resulting motion is also a bound motion.}

To analyse this situation, which is going to produce bound motions, we proceed as follows: 
We start the numerical integration by imposing 
the boundary condition that both centres of mass are at rest and located at the origin of the reference frame
${\bi q}_a(0)=\dot{\bi q}_a(0)=0$. For particle 2 we take the initial phase $\beta_2(0)=0$ 
and for $\beta_1$ we start with $\beta_1(0)=0$ and, will be increased step by step in one degree in the automatic process, 
up to reach the whole range of $2\pi$ radians. The boundary values of the 
variables ${\bi r}_a(0)$ and  $\dot{\bi r}_a(0)$, with the constraint $|\dot{r}_a(0)|=1$, 
are taken as the corresponding values compatible with these phases. 
The whole system is analysed in its centre-of-mass frame, so that for subsequent
boundary values these variables are restricted to ${\bi q}_1(0)=-{\bi q}_2(0)$
and $\dot{\bi q}_1(0)=-\dot{\bi q}_2(0)$. 
The automatic integration is performed in such a way
that when the two particles separate, i.e., when their centre-of-mass separation is above Compton's
wavelength, the integration stops and starts again with a new boundary value of the phase $\beta_1(0)$ of one degree
more, and the new values of the variables ${\bi r}_a(0)$ and  $\dot{\bi r}_a(0)$. 
If the two particles do not separate
we wait until the integration time corresponds to $10^6$ turns of the charges around their corresponding
centre-of-mass, stop the process, keep record of the phases and initial velocities, and start again
with new boundary values. This corresponds, in the case of electrons, 
to a bound state leaving during a time greater than $10^{-15}$ seconds. For some particular
boundary values, with opposite phases, 
we have left the program working during a whole week and the bound state prevails. This represents
a time of life of the bound state greater than $10^{-9}$ seconds. Leaving the computation program running 
for a year will only increase this lower bound in two orders of magnitude. The general feeling is that the bound states
are sufficiently stable, because even the possible numerical integration errors do not destroy the stability.
This process is repeated again and again by changing slightly the initial
values of the centre-of-mass variables ${\bi q}_a(0)$ and $\dot{\bi q}_a(0)$, in steps of $0.0001$ in these
dimensionless units and with $\beta_2(0)=0$, and the same procedure with $\beta_1(0)$, as above. 
To test the acuracy
of the integration method, we check every $10^3$ integration steps 
that the velocities of the charges of both particles remain of absolute value 1, within a numerical error smaller than
$10^{-20}$.

The whole process is repeated by changing the initial $\beta_2(0)$ phase to any other arbitrary value.
We are interested to see whether different results are produced depending on the values of 
the phase difference $\beta_2(0)-\beta_1(0)$
and of the centre-of-mass variables ${\bi q}_a(0)$ and $\dot{\bi q}_a(0)$. 
We collect all data which produce bound motions, and find the following results:
\begin{enumerate}
\item{The initial velocity of their centres of mass must be $|\dot{q}_a(0)|<0.01c$. 
Otherwise the bound motion is not stable and the two particles, after a few turns, go off.
}
\item{For each velocity $|\dot{q}_a(0)|<0.01c$ there is a range $\Delta$ of the pase $\beta_1(0)=\beta_2(0)+\pi\pm \Delta$ 
for which the bound motion is stable. The greater the centre-of-mass velocity of each particle 
the narrower this range, so that the bound motion
is more likely when the phases are opposite to each other.}
\item{We have found bound motions for an initial separation between the centres of mass up to $0.8$ times
Compton's wavelength, like the one depicted in figure \ref{fig:phaseL}, 
provided the above phases and velocities are kept within the mentioned ranges.}
\end{enumerate}

In figure \ref{fig:bound} we show the bound motion of both particles when their centres 
of mass are initially separated $q_{1x}=-q_{2x}=0.2\times$Compton's wavelength, $\dot{q}_{1x}=-\dot{q}_{2x}=0.008$
and $\dot{q}_{1y}=-\dot{q}_{2y}=0.001$, $\beta_2=0$ and $\beta_1=\pi$.
Now the force between the charges is repulsive but nevertheless, if the internal phases
$\beta_1$ and $\beta_2$ are opposite to each other, it becomes an atractive force between
their centres of mass in accordance to the mechanism shown in figure \ref{fig:phase} (b). 

This possibility of formation of low energy metastable 
bound pairs of particles of the same charge is not peculiar of this interaction
Lagrangian. By using the electromagnetic interaction or even the instantaneous 
Coulomb interaction between the charges
of two spinning Dirac particles we found in \cite{dyn} also this behaviour.
This bound motion is not destroyed by external electric fields and also by an external magnetic field along the
spin direction. Nevertheless, a transversal magnetic field destroys this bound pair system.

\cfigl{fig:bound}{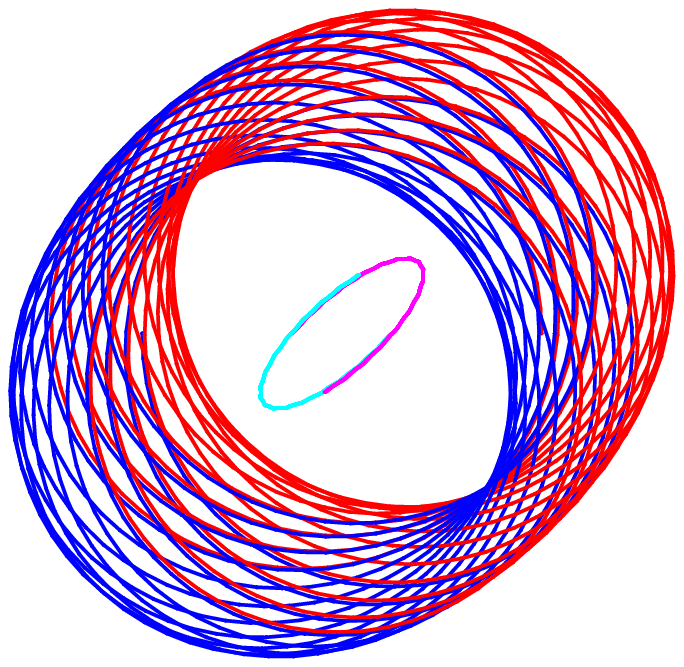}{Bound motion of the centres of mass and charge of two spinning particles 
with parallel spins and with a centre-of-mass velocity $v\simeq0.0082$, for an initial separation between the centres
of mass of $0.2\times$Compton's wavelength.}

When we make the average of the position ${\bi r}_a$ it becomes the centre-of-mass ${\bi q}_a$ and 
the repulsive force between the charges is also a repulsive force between the corresponding centres of mass
and therefore when we suppress the zitterbewegung spin content of the particles there is no possibility
of formation of bound pairs.

Although this result produces a classical mechanism for the formation of a spin 1 bound system 
from two equal charged fermions we must be careful about its conclusions. 
First, it is a classical description and although 
the range of energies
which produce this phenomenon is a wide one it does not mean that two electrons can reach that binding energy. 
This Dirac particle is a system of seven degrees of freedom: 3 represent the position ${\bi r}$, 
another 3 the orientation $\balpha$ and finally the phase $\beta$. 
If we accept the equipartition theorem for the energy, 
then for the maximum kinetic energy which produces a bound motion $mv^2/2=7\kappa T/2$, 
where $\kappa$ is Boltzmann's constant and
$v=0.01c$ the maximum velocity of the center of mass of each particle, then it means that a gas of polarized
electrons (like the conducting electrons in a quantum Hall effect) could form bound states up to a temperature 
below $T=8.47\times 10^5$K, which is a very high temperature.
In a second place, matter at this level behaves according to quantum mechanical rules and therefore
we must solve the corresponding quantum mechanical bound state to establish 
the proper energies and angular momenta at which these bound states would be stationary. 
This problem has not been solved yet, but
the existence of this classical possibility of formation of bound pairs 
justifies an effort in this direction. 
If the phases of the two particles are the same (or almost the same) there is no possibility
of formation of a bound state. The two fermions of the bound state have the same spin and energy.
They differ that their
phases and linear momenta are opposite to each other. Is this difference in the phase
a way to overcome at the classical level, the Pauli exclusion principle?

\ack{I thank my colleague J M Aguirregabiria for the use of his excellent
Dynamics Solver program \cite{JMA} with which the numerical computations of the examples
have been done. This work has been partially supported by 
Universidad del Pa\'{\i}s Vasco/Euskal Herriko Unibertsitatea grant  9/UPV00172.310-14456/2002.}


\end{document}